\documentclass[aps,prd,nofootinbib,twocolumn,superscriptaddress]{revtex4}

\usepackage{graphicx}
\usepackage{amsmath}
\usepackage{slashed}
\usepackage{color}
\usepackage[hidelinks]{hyperref}

\usepackage{amsfonts}
\usepackage{amssymb}
\usepackage{mathrsfs} 
\usepackage{cases} 
\usepackage{placeins} 
\usepackage{mathtools}

\newcommand{\phir}{\frac{\phi}{f}}
\newcommand{\mpl}{M_{P}}

\newcommand{\be}{\begin{equation}}
\newcommand{\ee}{\end{equation}}
\newcommand{\br}{\begin{eqnarray}}
\newcommand{\bea}{\begin{eqnarray}}
\newcommand{\eea}{\end{eqnarray}}
\newcommand{\er}{\end{eqnarray}}
\newcommand{\ba}{\begin{array}}
\newcommand{\ea}{\end{array}}
\newcommand{\bi}{\begin{itemize}}
\newcommand{\ei}{\end{itemize}}
\newcommand{\bn}{\begin{enumerate}}
\newcommand{\en}{\end{enumerate}}
\newcommand{\bc}{\begin{center}}
\newcommand{\ec}{\end{center}}

\newcommand{\Eq}[1]{Eq.~(\ref{#1})}
\newcommand{\rfn}[1]{(\ref{#1})}

\newcommand{\gsim}{\lower.7ex\hbox{$\;\stackrel{\textstyle>}{\sim}\;$}}
\newcommand{\lsim}{\lower.7ex\hbox{$\;\stackrel{\textstyle<}{\sim}\;$}}

\newcommand{\dg}{\dagger}   
\newcommand{\LG}{\mathcal{L}}

\newcommand{\deq}{\coloneqq}

\def\mysection#1{{\bf #1.} }

\begin{document}

\title{\Large Natural Relaxation}
\author{Luca Marzola}
\author{Martti Raidal}
%\email{martti.raidal@cern.ch}
\affiliation{National Institute of Chemical Physics and Biophysics, R\"avala 10, 10143 Tallinn, Estonia.}
\affiliation{Institute of Physics, University of Tartu, Ravila 14c, 50411 Tartu, Estonia. }

\begin{abstract}

Motivated by natural inflation, we propose a relaxation mechanism consistent with inflationary cosmology that explains the hierarchy between the electroweak scale and Planck scale. 
This scenario is based on a selection mechanism that identifies the low scale dynamics as the one that is screened from UV physics.
The scenario also predicts the near-criticality and metastability of the standard model vacuum state, explaining the Higgs boson mass observed at the LHC. 
Once Majorana right-handed neutrinos are introduced to provide a viable reheating channel, our framework yields a corresponding mass scale that allows for the seesaw mechanism as well as for standard thermal leptogenesis. We  argue that considering singlet scalar dark matter extensions of the proposed scenario could solve the vacuum stability problem and discuss how the cosmological constant problem is possibly addressed.

\end{abstract}

%\date{}
%\makeindex

\maketitle

%\maketitle

%===============================================================================
% BODY
%===============================================================================
%-------------------------------------------------------------------------------
\section{Introduction} % (fold)
\label{sec:Introduction}
%-------------------------------------------------------------------------------

Despite the existence of the standard model (SM) Higgs boson has been experimentally verified~\cite{Chatrchyan:2012xdj,Aad:2012tfa}, the mechanism that protects its mass from the influence of a potential high-energy cut-off remains still mysterious. Every new and technically consistent view on the matter, know in literature as the hierarchy problem, must then be pursued and scrutinised. For instance, the recently proposed relaxion mechanism~\cite{Graham:2015cka} suggests a conceptually new approach to tackle this issue. More in detail, the scenario is based on the interplay between the Higgs boson, that receives a large quadratically divergent mass correction from the high scale cut-off scale $\Lambda$, and a new periodic axion-like field which is instead protected against the same corrections by a shift symmetry. With this setup, the relaxion mechanism exploits 
$(i)$ the dynamical relaxation of the Higgs boson mass term \cite{Dvali:2003br,Dvali:2004tma} due here to the dynamics of the axion-like field and 
$(ii)$ a feedback effect from the electroweak symmetry breaking which stops the relaxation process at the right scale
to solve the hierarchy problem. Besides the axion-like field, the scenario does not introduce any new particle around the electroweak scale to explain the naturalness of the latter.

The relaxion scenario is currently the subject of a thorough investigation. Unfortunately, for the required huge number of $e$-folds and very low inflation scale, the model presented in Ref.~\cite{Graham:2015cka} seems to be at odds with cosmology. No inflation scenario consistent with both the relaxion mechanism and the Planck/BICEP2/Keck data~\cite{Ade:2015lrj} is presently known. Another possible shortcoming of this solution is that only a cut-off scale $\Lambda$ much below any supposed new physics scale, such as the GUT, string or Planck scales, can be relaxed to the desired values. In addition to that, the mechanism has been criticised for introducing a fine-tuning in a different sector that is more severe than the one required by the hierarchy problem within the SM~\cite{Jaeckel:2015txa}.
The proposed modifications of this scenario~\cite{Espinosa:2015eda,Hardy:2015laa,Patil:2015oxa,Gupta:2015uea,Batell:2015fma,Antipin:2015jia,Matsedonskyi:2015xta,Kobakhidze:2015jya} 
mostly try to improve on one or several of these issues, leaving aside the matter of a possible ultraviolet (UV) completion of the theory.

In this work we present a new electroweak scale relaxation scenario that is consistent with inflationary cosmology and that is able to relax scales as large as the Planck scale. Our mechanism draws from the periodic scalar potential of {\it natural inflation}~\cite{Freese:1990rb,Adams:1992bn} and allows us to identify the field that drives the relaxation mechanism with the inflaton. Since, to date, natural inflation is consistent with the cosmological measurements~\cite{Ade:2015lrj}, our scenario is automatically in line with the corresponding constraints. Furthermore, our mechanism shares the same UV-completion as the original natural inflation scheme --  the string theory -- that supports the existence of many scalar fields. 

The basis of our construction is in a selection mechanism that determines which scalar fields, among the many candidates in string theory, are screened from the UV physics located at the cutoff scale. The screened scalars then become naturally much lighter than the latter, while the remaining particles remain at the cutoff scale. As we will show, our selection mechanism is based on the interactions between the scalar fields and the inflaton. Only the scalars that couple to this particle in a minimal will contribute to the low-energy dynamics of the SM.
As in the original natural inflation theory, the shape of the inflaton potential and, more in general, of its interactions, is dictated by the fact that the inflaton mimics a pseudo-Nambu-Goldstone boson~\cite{Freese:1993bc}. While the inflaton slowly rolls towards the minimum, the potential of a screened scalar field, the SM Higgs, is relaxed into the desired form. No additional feedback mechanism is needed to stop the rolling since the potential has a well-defined minimum around which the inflaton field oscillates and consequently reheats the Universe.
The essential ingredient of this {\it natural relaxation} mechanism is the shift symmetry of the periodic inflaton potential. The inflation dynamics selects one of the many minima, breaking the shift symmetry and consequently inducing a naturally small and non-vanishing electroweak scale.

The natural relaxation mechanism has interesting predictions that are consistent with phenomenology. 
Firstly, it predicts an almost vanishing Higgs boson quartic coupling at the inflation scale. This explains the mysterious 
near-criticality of the SM vacuum~\cite{Froggatt:1995rt} and predicts the observed Higgs boson mass.
In our framework, the Higgs boson SM couplings generate at one-loop a small but non-vanishing effective Higgs quartic coupling.
In this regard, we compute the renormalization group improved one-loop effective potential for the SM Higgs boson and show that the 
effective quartic is small and {\it negative}, explaining the metastability of the SM vacuum~\cite{Degrassi:2012ry,Buttazzo:2013uya}.
Extending our framework with scalar singlet dark matter~\cite{Silveira:1985rk} could furthermore solve the vacuum stability problem in a way consistent with the 
observed Higgs boson properties~\cite{Kadastik:2011aa}.
If the SM is extended instead with right-handed Majorana neutrinos, as we assume here, the shift symmetry breaking induces 
a right-handed neutrino mass scale that is consistent with the seesaw explanation of the smallness of the active neutrino masses~\cite{Minkowski:1977sc} and 
with the baryon asymmetry of the Universe via leptogenesis~\cite{Fukugita:1986hr}.
The natural relaxation mechanism may also help to relax the cosmological constant~\cite{Weinberg:1988cp}. 
Due to the shift symmetry breaking, however, a cosmological constant at least of order of the fourth power of the electroweak scale is necessarily generated within our framework. We argue that the most natural dynamical mechanisms for addressing the remaining problem are based on particle creation by de~Sitter vacuum during inflation, which could induce the decay of the relaxed cosmological constant~\cite{Myhrvold:1983hx,Mottola:1984ar,Antoniadis:1985pj,Freese:1986dd,Ford:1987de,Tsamis:1996qq}.

% section Introduction (end)
%-------------------------------------------------------------------------------
\section{Natural inflation and relaxation of the electroweak scale} % (fold)
\label{sec:The model}
%-------------------------------------------------------------------------------

We consider a model in which the SM particle content is augmented by three generations of singlet right-handed neutrino fields $N_i$, whose mass is dynamically generated. The scalar sector of the theory contains, amongst other particles, the SM Higgs boson doublet $H$ and a real scalar field $\phi$. The latter is a pseudo-Nambu-Goldstone boson, serves as an inflaton, and plays a crucial role in our relaxation mechanism. 
Our setup is meant to mimic an effective theory specified at the high-energy cutoff  resulting from a possible incarnation of string theory that introduces for the field $\phi$ a periodic scalar potential invariant under a discrete shift symmetry $\phi \to \phi + 2n\pi f$, $n \in \mathbb{N}$:
\bea
	\label{eq:lagnr}
	V &=&
	-\Lambda^2 \left(1 + \cos\left(\phir\right)\right) H^\dg H  \\
	& +& \lambda \left(1 + \cos\left(\phir\right)\right)\left(H^\dg H\right)^2 
	 +\Lambda^4\left(1 + \cos\left(\phir\right)\right). \nonumber
\eea
We expect that all the mass scales, as well as the scale $f$ where the periodic potential is generated, be of order of the cut-off scale ${\cal O}(\Lambda)$.
The last term in \Eq{eq:lagnr} is the usual potential of natural inflation~\cite{Freese:1990rb,Adams:1992bn}, which is one of the few inflation scenarios consistent, to date, with both the Planck and BICEP2 results~\cite{Ade:2015lrj}. The remaning terms generalise the characteristic pseudo-Goldstone interaction to the possible dimension four operators that involve the Higgs doublet and the periodic potential. Notice that corresponding terms for the inflaton $\phi$ are forbidden by the shift symmetry. 

According to the proposed paradigm, if a scalar particle of the theory couples to the inflaton minimally, according to Eq.~\rfn{eq:lagnr}, the same dynamics that drives the inflation screens the particle from the effects of UV physics, generating a low-energy potential. The remaining scalar fields whose potentials do not match Eq.~\rfn{eq:lagnr},
which number is expected to be large in the string theory, are naturally heavy and form a decoupled sector. Thus, only minimally coupled scalars will play a role in the low-scale dynamics of the SM and, for simplicity, we will assume the existence of only one such scalar field -- the Higgs boson.

Focusing now on the Higgs boson properties, amongst the terms in the presented potential we can distinguish an effective mass term $\mu_{\text{eff}}^2 H^\dg H$ 
\begin{equation}
	\label{eq:mu_eff}
	-\mu_{\text{eff}}^2 = \Lambda^2 \left(1 + \cos\left(\phir\right)\right),
\end{equation}
which depends on the evolution of the inflaton field $\phi$. The latter is driven by the inflaton potential that dominates the early Universe before reheating.
During natural inflation, the inflaton slowly rolls towards the minimum of its potential and then, oscillating around this point, reheats the Universe. Since all the terms in \Eq{eq:lagnr} depend on the evolution of the inflaton field in the same way, at the classical level, after inflation all the terms are dynamically driven to vanishing values. 

This is the essence of the natural relaxation mechanism. There are important differences by which our mechanism can be distinguished from the original relaxion model~\cite{Graham:2015cka}. The scalar potential has well defined degenerate minima and the inflation dynamics is responsible for selecting one of them, establishing the physical vacuum state. In the natural relaxation mechanism the inflaton falls into one of the minima satisfying the slow-roll conditions, and no back-reaction mechanism from the electroweak symmetry breaking is required to stop the field dynamics. 
Since the field that dynamically relaxes the SM Higgs boson mass is the inflaton of natural inflation, the accordance with cosmology is guaranteed by the robustness of the latter. 

The dynamics that drives $\phi$ to a minimum clearly breaks the shift symmetry by selecting \emph{one} minimum. 
This breaking introduces {\it naturally small} effects that generate, at quantum level, deviations from the classical configuration for which $\left(1 + \cos\left(\phi/f\right)\right) = 0$ and, consequently, a small but non-vanishing Higgs boson mass term. 
The smallness of the electroweak scale compared to $\Lambda$, being here a consequence of the shift symmetry breaking, is technically natural.

To model these quantum effects, we introduce the following operators that break the shift symmetry explicitly,
\begin{equation}
	\label{eq:lagrnr_sb}
	\LG \supset \LG_{\text{sb}} = - \sum_{i,j=1}^3 y^{ij} \phi \bar N_i^c N_j   + \kappa \phi (H^\dg H)  + {\cal O}(\phi^n) + {h.c.},
\end{equation}
where $y$  denote the Yukawa couplings of the right-handed neutrinos, $\kappa$ is a coupling with dimension of a mass 
and $ {\cal O}(\phi^n)$ denotes implicitly all the terms of higher order in $\phi$. We assume the latter be subdominant because of higher order also in the shift symmetry breaking parameters. Restricting ourselves to interactions linear in $\phi$, the particle content of our framework admits exclusively the terms indicated in Eq.~\eqref{eq:lagrnr_sb}.
Notice that on general grounds such operators \rfn{eq:lagrnr_sb} are required by thermal reheating, which in this framework proceeds through the decay of the field $\phi$.
More in general, the presence of such interactions is required in every natural inflation model and does not depend on the specifics of our construction. 
In our scheme, however, these operators play a fundamental part as they could prevent the exact relaxation of the terms in Eq.~\eqref{eq:lagnr} and consequently generate the electroweak scale. 

%-------------------------------------------------------------------------------
\section{The electroweak symmetry breaking} % (fold)
\label{sec:The role of quantum corrections}
%-------------------------------------------------------------------------------

Our framework proposes two viable mechanisms leading to the SM electroweak symmetry breaking, depending on which of the two terms in \Eq{eq:lagrnr_sb} is dominating.
If the interaction with the RH neutrino is negligible, the SM Higgs boson mass term $\mu^2_{\text{SM}}H^\dg H$ is generated at tree level by the $\kappa \phi (H^\dg H)$ term.  At the  minimum of the potential, to a good approximation $\phi \approx \pi f$, implying
\bea
\mu_{\text{SM}}^2\approx \pi \kappa f.
\label{mu}
\eea
 If $\kappa<0$, the Higgs mass term is negative and the electroweak symmetry breaking can be triggered.  Numerically, for $f\sim \Lambda\sim M_P$, reproducing the desired hierarchy $\mathcal{O}(-\mu^2_{\text{SM}}/\Lambda^2)\simeq 10^{-34}$ imposes $\left|\kappa\right|\sim v^2/M_P\sim 10^{-6}$~eV. This value is in agreement with the general expectation that symmetry breaking parameters are naturally small.
  
Notice, however, that the shift symmetry breaking term $\kappa \phi (H^\dg H)$ induces new contributions to the scalar potential which can be quantified by minimising the `improved' potential with respect to $\phi$:
\begin{equation}
	V_1^{\text{imp}} = \Lambda^4\left(1 + \cos\left(\phir\right)\right) + \kappa \phi (H^\dg H).
	\label{eq:Vimp1}
\end{equation}
We set the Higgs field value at $H= v/\sqrt{2}$ and expand the inflaton field around the minimum $\phi=\phi_0$ of its potential to compute the deviation
$\delta\deq \pi - \phi_0/f$ from the tree-level minimum $\pi f$. We obtain
\begin{equation}
	\delta = \frac{\kappa  f v^2}{2\Lambda^4}\approx{\cal O}(v^4/\Lambda^4)\sim 10^{-68}.
\end{equation}
Consequently, the contribution \rfn{eq:mu_eff} to the Higgs mass term is completely negligible compared to
the one of \rfn{mu} as the displacement originated from the quantum correction has no significant effect. The term $\kappa \phi (H^\dg H)$ alone is then responsible for the electroweak symmetry breaking.

Differently, if the Yukawa term $y\phi NN $ dominates, the right-handed neutrino Yukawa couplings generate at one loop level a mass term for the inflaton and, to the lowest order in symmetry breaking, a new quadratic component of the inflaton potential.  
Therefore, in this case we minimise the following improved potential  
\begin{equation}
	V_2^{\text{imp}} = \Lambda^4\left(1 + \cos\left(\phir\right)\right) + m^2 \phi^2,
	\label{eq:Vimp}
\end{equation}
where $m$ is the mentioned mass term generated at one-loop level by the interaction in Eq.~\eqref{eq:lagrnr_sb}. 
By applying the usual minimization procedure and expanding around the classical minimum $\phi_0$ as well as in the small parameter $2m^2f^2/\Lambda^4 \ll 1$, we then find the deviation $\delta\deq \pi - \phi_0/f$,
\begin{equation}
	\delta = \frac{2\pi m^2 f^2}{\Lambda^4}.
\end{equation}
Requiring now that this effect prevent the exact cancellation of the effective Higgs mass  yields $\left(1 + \cos\left(\phi/f\right)\right) \approx 10^{-34}.$ Consequently $\delta \approx O(10^{-17})$ and we must have $mf \approx 10^{-9} \Lambda^2$. If we regard the quadratic divergences brought to scalar masses by loop correction as physical, which is the assumption behind the large Higgs mass in Eq.~\eqref{eq:lagnr}, then $m \approx y \Lambda$ and therefore $y f \approx 10^{-9} \Lambda$. Then, as we expect $f\sim \Lambda$, this construction is perfectly consistent with the requirement that the shift-symmetry breaking couplings $y$ in Eq.~\eqref{eq:lagrnr_sb} be small. Furthermore, given that the field $\phi$ acquires a vacuum expectation value $\phi \approx f\pi$, the fermions $N_i$ in Eq.~\eqref{eq:lagrnr_sb} acquire a typical right-handed neutrino mass scale $M_N \approx y f \pi \approx 10^{-9} \Lambda\approx 10^{10}$ GeV, having identified here $\Lambda \equiv \mpl$. Once the type-I seesaw mechanism~\cite{Minkowski:1977sc} is considered, this scale is remarkably in agreement with the smallness of the active neutrino masses indicated by the oscillation experiments, as well as with the measured baryon asymmetry of the Universe which can be here generated via the standard thermal leptogenesis~\cite{Fukugita:1986hr}.

We emphasize that despite relying on the crucial effects of quantum correction, the above scenario is conceptually different from the Coleman-Weinberg mechanism~\cite{Coleman:1973jx} proposed for the generation of the one-loop effective potential of the Higgs boson.
Here, the symmetry breaking term  $y\phi NN $ generates a loop correction to the inflaton potential that induces a shift in the inflaton field minimum, which results in a non-vanishing contribution to the Higgs mass via \Eq{eq:mu_eff}.

%-------------------------------------------------------------------------------
\section{The Higgs quartic coupling at high energies} % (fold)
\label{sub:The Higgs quartic coupling}
%-------------------------------------------------------------------------------

Because of the relaxation mechanism, at tree level also the Higgs quartic coupling naturally vanishes at high energies. 
As we will now show, this is a good starting point to address the observed criticality of the SM vacuum at high scales. The 
Higgs doublet clearly interacts with the remaining SM matter fields and gauge bosons and, as a result, the Higgs boson quartic coupling receives additional contributions induced at the loop level by the same SM particle content.
To estimate this effect, we compute the renormalization group improved effective potential for the Higgs boson at one-loop level. The fourth derivative
of the latter evaluated at large field values then returns the effective SM quartic coupling $\lambda_{\text{eff}}$.

In the $\overline{MS}$ renormalization scheme the one loop Higgs effective potential is given by~\cite{Sher:1988mj}
\begin{equation}
\label{V_eff}
	V_{\text{eff}}(H) = \sum_\alpha\frac{N_\alpha M_\alpha^4}{64 \pi^2}
	\left(\log\left(\frac{M_\alpha^2}{Q^2}\right)-C_\alpha\right),
\end{equation}
where $\alpha\in\{Z, W, t\}$, $N_Z = 3$, $N_W = 6$, $N_t = -12$, and 
\bea
\label{Mt}
	M_Z^2 &=& \left(g^2 + g^{\prime\,2}\right)H^\dg H /4
	, \qquad
	M_W^2 = g^2 H^\dg H/4
	, \nonumber \\
	M_t^2 &=& y_t^2 H^\dg H /2,
\eea
and $C_\alpha=3/2$ for fermions and $5/6$ for gauge bosons. Because the tree level Higgs quartic coupling is negligibly small, we have suppressed in \Eq{V_eff} the 
Higgs and Goldstone boson contributions. In our computation of the improved effective potential we use the two-loop SM 
renormalization group equations~\cite{Machacek:1983fi,Machacek:1984zw} to
evaluate the SM couplings at the field dependent renormalisation scale $Q=M_t$ given by \rfn{Mt}.
We use the following NLO input values at the top mass scale~\cite{Buttazzo:2013uya}:
$g_Y=0.35940$, $g_2=0.64754$, $g_3 = 1.666$, $y_t=0.95113$, $\lambda=0.12774$.

\begin{figure}[t]
   \centering
     \includegraphics[width=.48\textwidth]{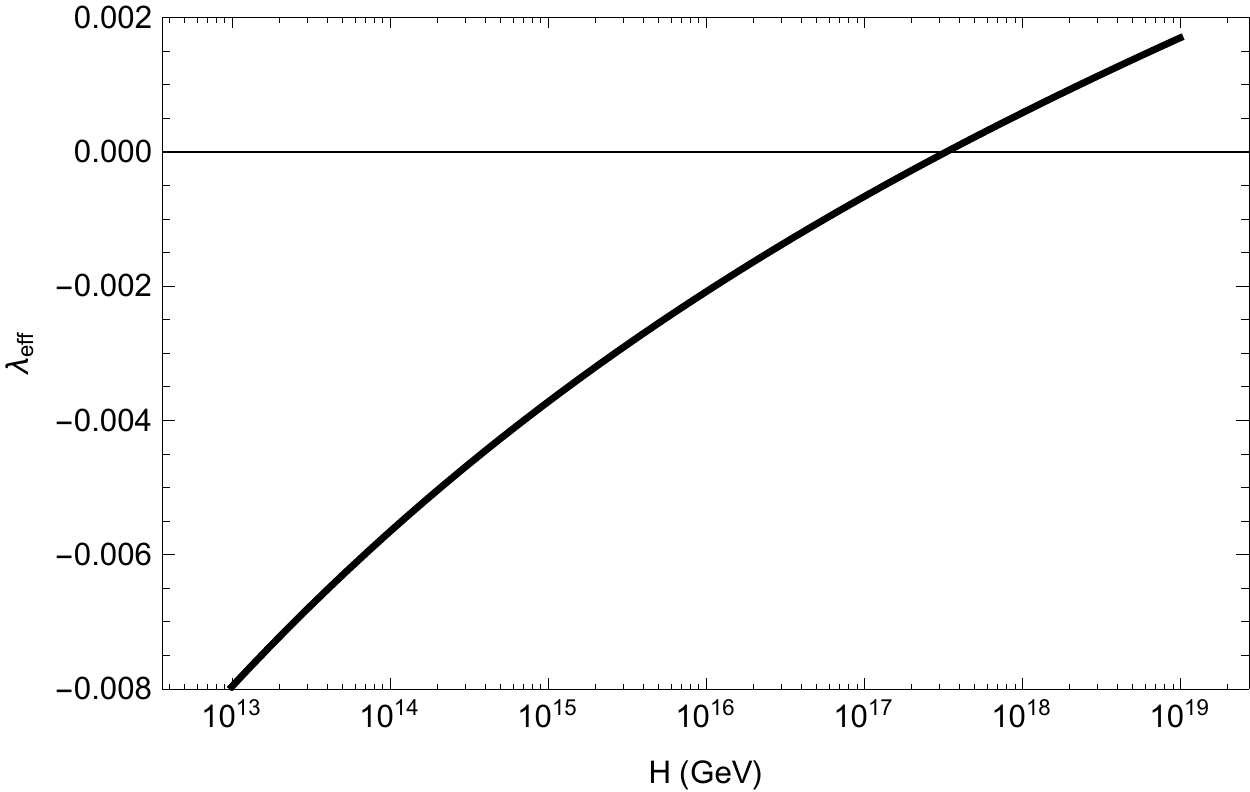}
   \caption{The generated effective quartic coupling of the SM Higgs field as a function of the field value.}
   \label{fig:lambda}
 \end{figure}

The resulting effective Higgs quartic coupling is plotted in Fig.~\ref{fig:lambda} as a function of the field value.
During inflation, the Hubble parameter is of order $H\sim10^{14}$~GeV. While the Higgs field fluctuates, its average value is of the same order of the Hubble parameter and, therefore, we have to evaluate the coupling at the corresponding high-energy scale.
Because the behaviour of the SM quantum corrections is such that the gauge bosons and top quark contributions tend to cancel, $\lambda_{\text{eff}}$ is positive above
$5\times 10^{17}$~GeV and negative for smaller field values. Therefore, our scenario effectively predicts that the SM Higgs quartic coupling 
after inflation is $\lambda_{\text{eff}}\approx -0.006,$ consistently with the metastability of the SM vacuum.

%-------------------------------------------------------------------------------
\section{Extensions of the model and vacuum stability}
%-------------------------------------------------------------------------------

So far we focused on the SM and showed how its the measured properties emerge in our framework, included the instability of the vacuum state. Although phenomenologically acceptable, the metastability of our Universe is a disturbing idea. Immediately after the LHC determined the Higgs boson mass, 
it was pointed out in Ref.~\cite{Kadastik:2011aa} that a simple singlet scalar extension of the SM could solve the vacuum stability problem
due to new, positive, contribution to the renormalization group equations of the Higgs quartic coupling emerging from a portal coupling $\lambda_{SH}S^2H^2$~\cite{Gabrielli:2013hma}.

For this solution to be implemented within our framework, the new scalar potential terms must be of the form contained in \rfn{eq:lagnr} and the associated mass scales will be relaxed in analogy to the Higgs one. It could then be argued that the same relaxation mechanism could drive the portal coupling $\lambda_{SH}$ to negligible values, therefore, invalidating the mentioned solution. However, notice that introducing a new light scalar singlet $S$ in the model allows for several new 
shift symmetry breaking terms as
\bea
\label{sb2}
\lambda_1 \phi SSS, \qquad \lambda_2 \phi S |H|^2, 
\eea
which add to those of \Eq{eq:lagrnr_sb}. After the inflaton is driven to its final field value, $\phi_0\approx \pi f,$ the two terms in \Eq{sb2}
produce a tree level effective interaction $\lambda_{SH}^{\text{eff}}S^2H^2$.
Once this is generated, the renormalization group improved effective potential \Eq{V_eff} receives a new, {\it positive} contribution with $N_S=1$ and 
$M^2_S=\lambda_{SH}^{\text{eff}} H^2$~\cite{Espinosa:2007qk}, which can easily overcome the SM contribution yielding a positive $V_{\text{eff}}$ at high energies.

%-------------------------------------------------------------------------------
\section{The cosmological constant problem} % (fold)
\label{sec:The cosmological constant problem}
%-------------------------------------------------------------------------------

The bare cosmological constant in our scenario is naturally as large as the cut-off scale of the theory, $\Lambda \sim M_P$, because of the quantum corrections induced by the latter. However, when the inflaton relaxes the potential \rfn{eq:lagnr} to $V=0,$ the cosmological constant problem~\cite{Weinberg:1988cp} should be technically solved up to corrections due to physics below the electroweak scale. 

Unfortunately the picture is more complicated than that. Due to the shift symmetry breaking, new terms are generated in the scalar potential, as explicitly demonstrated by
\Eq{eq:Vimp1} and \Eq{eq:Vimp}. In the first case, where the mass of the Higgs boson is generated at tree level, we can expect a new contribution to the cosmological constant of order ${\cal O}(v^4)$. With respect to the SM, the situation is then improved to the level of the cosmological constant problem within supersymmetric models. 

There are now two possible solutions. The first, less appealing possibility consists in including all the allowed low-energy contributions and fine-tuning the value of the obtained scalar potential according to the phenomenological requirements. Alternatively, we could assume that a further dynamical mechanism reduces the remaining vacuum energy down to the observed value. The latter possibility is natural within our framework since the known mechanisms rely on the decay of de~Sitter space vacuum~\cite{Myhrvold:1983hx,Mottola:1984ar,Antoniadis:1985pj,Freese:1986dd,Ford:1987de,Tsamis:1996qq}, which can take place during the inflation era. While the back-reaction effects in these theories have the correct sign to reduce the bare vacuum energy, it is clear that no definitive claim can be made before having performed a full computation within the right theory of quantum gravity. 

% section The cosmological constant problem (end)

%-------------------------------------------------------------------------------
\section{Conclusions} % (fold)
\label{sec:Conclusions}
%-------------------------------------------------------------------------------

We present the natural relaxation mechanism for the generation of the electroweak scale. 
Our mechanism is based on the identification of the field that drives the relaxation of the SM Higgs potential with the inflaton of natural inflation, a pseudo-Nambu-Goldstone boson whose dynamics is characterised by a periodic potential. The proposed framework explains the hierarchy between the electroweak and Planck scales in a way consistent with inflationary cosmology and predicts both the near-criticality and meta-stability of the SM vacuum. We advocate that the latter problem can be solved in the present framework by adding a singlet scalar dark matter candidate, exactly as in the SM. The scenario we propose also generates a right-handed neutrino mass scale that allows for the seesaw type-I mechanism and leptogenesis. 
On top of that, we have argued that in our framework the cosmological constant problem is alleviated by many orders of magnitude. A complete solution of the problem might involve additional dynamical mechanism such as the decay of de~Sitter vacuum during the inflation era. Finally, we remark that the proposed solution requires that light scalar particles of the theory, such as the Higgs boson and possibly Dark Matter, couple to the inflaton in a specific manner, yielding a distinguishing UV completion of the model that generally predicts a heavy scalar sector at the cutoff energy. The discovery at the LHC of competing mechanisms for the stabilization of the electroweak scale, such as supersymmetry for instance, would indeed severely disfavour our mechanism.

% section Conclusions (end)
%-------------------------------------------------------------------------------
%\subsection*{Acknowledgement} % (fold)
%\label{sec:Acknowledgements}
%-------------------------------------------------------------------------------
\mysection{Acknowledgement} The authors thank Kristjan Kannike, Antonio Racioppi and Alessandro Strumia for useful discussions and for their help with numerical codes. 
This work was supported by the grants IUT23-6, PUTJD110, CERN+ and by EU through the ERDF CoE program.

% section Aknowledgements (end)
%\newpage

%\bibliographystyle{unsrt}
%\addcontentsline{toc}{chapter}{References}
\bibliography{bib}

\end{document}